\documentstyle[12pt,epsf]{article}

\newcommand{\be}{\begin{equation}}
\newcommand{\ee}{\end{equation}}
\newcount\sectionnumber
\def\sect
{\global\equationnumber=0
\global\advance\sectionnumber by 1
\the\sectionnumber . }
\newcount\equationnumber
\def \num
{\eqno{\global\advance\equationnumber by 1
\left(\the\sectionnumber .\the\equationnumber \right)}}

\def\boxit#1{\vbox{\hrule\hrule\hrule\hbox{\vrule\vrule\vrule\kern3pt\vbox{\kern3pt#1\kern3pt}\kern3pt\vrule\vrule\vrule}\hrule\hrule\hrule}}

\def\wave{ \vcenter{\vbox{\hrule height.3pt \hbox{\vrule width.3pt height9pt
\kern9pt \vrule width.3pt} \hrule height.3pt}} \, }
\def\comma{ \hspace{2mm}, }
\def\period{ \hspace{2mm}. }
\def\s{ \hspace{1mm} }
\def\b{ \hspace{2mm} }
\def\B{ \hspace{4mm} }
\def\EXP{ {\rm e} }
\def\Del{ \nabla }
\def\di{ \partial }
\def\de{ {\rm d} }
\def\half{ \frac{1}{2} }
\def\quarter{ \frac{1}{4} }
\def\minus{ \mbox - \, }
\def\const{ 8\,\pi }

\def\dix{ \partial_x }

\def\dixx{ \partial_{xx} }

\begin{document}
\pagenumbering{roman}
\title{Mass inflation in $(1+1)$-dimensional \\ dilaton gravity}

\author{J. S. F. Chan \\
Department of Applied Mathematics, \\
University of Waterloo, \\
Waterloo, Ontario, Canada \\
\vspace{10pt}
N2L 3G1 \\
R. B. Mann\\
Department of Physics, \\
University of Waterloo, \\
Waterloo, Ontario, Canada \\
N2L 3G1}

\date{}
\maketitle

\begin{abstract}
  We investigate the phenomenon of mass inflation in two-dimen\-sional
  dilaton theories of gravity.  We consider two distinct black hole
  spacetimes and construct the mass-inflation solution for each. Our
  analysis is extended to include multi-horizon spacetimes. We find that
  the mass function diverges in a manner quantitatively similar to its
  four-dimensional counterpart.
\end{abstract}

\pagenumbering{arabic}

\bigskip

\section{Introduction} \label{Sct1}
  Lower dimensional gravity continues to attract the attention of
  theorists, in large part because much of the conceptually
  interesting features of their $(3+1)$-dimensional counterparts are
  retained whilst the corresponding calculations involved are much simpler.
  Investigations of $(1+1)$-dimensional gravity \cite{lower} have proven to
  be particularly rewarding, providing further insight into black hole
  solutions \cite{Mann1}, \cite{Callan}, black hole radiation
  \cite{Morsink}, cosmology \cite{Ross}, singularities \cite{Dan,Chan} and
  quantum gravity \cite{Banks}.

  It was first pointed out by Penrose \cite{Penrose} that the Cauchy
  horizon of the Kerr-Newman solution of $(3+1)$-dimensional
  general relativity is unstable due to the
  infinite blueshift of the ingoing radiation. This pathology of
  the Cauchy horizon raises a question of whether the Kerr-Newman
  solution can be analytically extented to some other asymptotically
  flat universes. Much more recently Poisson and Israel \cite{Poisson}
  made important progress on this problem by showing that the Cauchy
  horizon of the Reissner-Nordstr\"{o}m solution forbids any evolution of
  spacetime beyond this horizon.
  They came to this conclusion by noticing that the mass
  parameter inside the black hole becomes unbounded due to the presence of
  ingoing and backscattered outgoing radiation. As the causal structure of
  the Kerr-Newman spacetime is similar to that of the
  Reissner-Nordstr\"{o}m case, it is believed that the Cauchy horizon in
  the more general Kerr-Newman case forms a similar obstruction to the
  evolution of the spacetime. Ori \cite{Ori} subsequently
  confirmed this mass inflation phenomenon by
  constructing an exact solution of the Einstein-Maxwell equations in a
  simpler model. However, he argued that this mass inflation singularity is
  so weak that its tidal forces do not necessarily destroy any physical
  objects. This extensibility problem has initiated some controversy among
  physicists in the field \cite{Herman,Bonnano}.

  Recently Husain \cite{Husain} showed that the mass inflation
  phenomenon also occurs in $(2+1)$-dimensional spacetime. In this
  paper, we shall show that the same phenomenon has analogous
  partners in $(1+1)$-dimensional gravitational theories. Specifically,
  we consider a construction of the Ori model in dilaton gravity theories
  in $(1+1)$ dimensions in which
  the background spacetimes possess more than two horizons.
  Even though these multi-horizon spacetimes have unusual
  properties, we find that the mass parameters generically
  inflate behind the outer horizon.

  We consider a non-critical string-inspired dilaton theory of gravity with
  the action of the form \cite{Nappi}
  \begin{eqnarray}
    S & = &
    \int \de^2x\,\sqrt{\minus g}\,\EXP^{\minus 2\,\phi}\,\left[\,R +
\gamma\,\left(\,\Del \phi\,\right)^2 - \quarter\,F_{\beta \sigma}\,F^{\beta
\sigma} + Q^2 + \sigma\,\EXP^{2\,\phi}\,\right] \nonumber \\ &&
    + \s \int \de^2x\,\sqrt{\minus g}\,\left[\,\sum^{k}_{n=2}
a_n\,\EXP^{2\,(n-1)\,\phi} - \const\,{\cal L}_M\,\right] \comma \label{S1E1}
  \end{eqnarray}
  where $R$ is the Ricci scalar, $\gamma$, $Q$, $\sigma$ and $a_n$'s are
  constants, $F_{\beta \sigma}$ is the Maxwell tensor and ${\cal L}_M$ is
  the matter Lagrangian. A choice of the values  of the parameters
  $\gamma$, $Q$, $\sigma$, and the $a_n$ is tantamount to a particular
  choice of theory.  We shall consider two distinct choices, each of which
  yields field equations whose solutions correspond to multi-horizon
  spacetimes.


\bigskip

\section{The First Solution} \label{Sct2}
  When $\gamma = 4$, $\sigma = 0$ and the parameter $Q$ is set to be
  a positive constant, the action (\ref{S1E1}) gives the field equations
  \be
    0 \b = \b
    \Del_\beta \left(\,\EXP^{\minus 2\,\phi}\,F_{\mu}{}^{\beta}\,\right) \comma
\label{S2E1A}
  \ee
  \be
    0 \b = \b
    R + 4\,\Del^2 \phi - 4\,\left(\,\Del \phi\,\right)^2
    - \quarter\,F^2 + Q^2
    - \sum^{k}_{n=2} a_n\,(n-1)\,\EXP^{2\,n\,\phi} \comma \label{S2E1B}
  \ee
  \begin{eqnarray}
    \const\,\EXP^{2\,\phi}\,T_{\mu \nu} & = &
    2\,\Del_\mu \Del_\nu \phi - 2\,\Del^2 \phi\,g_{\mu \nu}
    + 2\,\left(\,\Del \phi\,\right)^2\,g_{\mu \nu}
    - \half\,Q^2\,g_{\mu \nu} \nonumber \\ &&
    - \s \half\,\left[\,F_{\mu \beta}\,F_{\nu}{}^{\beta} -
\quarter\,F^2\,g_{\mu \nu}\,\right]
    - \s\half\,g_{\mu \nu}\,\sum^{k}_{n=2} a_n\,\EXP^{2\,n\,\phi} \period
\label{S2E1C}
  \end{eqnarray}
  When these equations are expressed in Eddington-Finkelstein
  ingoing coordinates
  \begin{eqnarray}
    \de s^2 & = & 2\,\de x\,\de v - \alpha(x,v)\,\de v^2 \comma \label{S2E2}
  \end{eqnarray}
  they have a solution
  \begin{eqnarray}
    f & = & q\,\EXP^{2\,\phi} \comma \label{S2E3A} \\
    \phi & = & \minus \frac{Q}{2}\,(x - x_o) \comma \label{S2E3B} \\
    \alpha & = &
    \frac{2}{Q}\,\left[\,{\cal M}(x) - m(v)\,\right]\,\EXP^{2\,\phi} \comma
\label{S2E3C} \\
    {\cal M}(x) & = &
    \frac{Q}{2}\,\EXP^{\minus 2\,\phi(x)}\,
    \left[\,1 + \frac{q^2}{2\,Q^2}\,\EXP^{4\,\phi(x)}
    - \frac{1}{Q^2}\,\sum^{k}_{n=2} \frac{a_n}{n-1}\,
    \EXP^{2\,n\,\phi(x)}\,\right] \label{S2E3D}
  \end{eqnarray}
  where the stress-energy tensor is
  \begin{eqnarray}
    T_{\mu \nu} & = & \rho(v)\,l_{\mu}\,l_{\nu} \period \label{S2E4}
  \end{eqnarray}
  The terms $x_o$ and $q$ are integration constants and $l_{\mu}$
  is defined as $l_{\mu} := \minus \di_\mu v$. The function $m(v)$
  satisfies the differential equation
  \begin{eqnarray}
    \const\,\rho(v) & = & m'(v) \period \label{S2E5}
  \end{eqnarray}
  This solution is asymptotically flat as $x \rightarrow \infty$ but
  is singular when $x \rightarrow \minus \infty$. One can show that
  the function $m(v)$ is the ADM mass \cite{Mann2} of the spacetime
  and this solution represents a black hole which is irradiated by
  an influx of null radiation.

  Consider the matching of two patches of solution
  (\ref{S2E3A})--(\ref{S2E3D}) along an outgoing null line S as shown in
  Figure~\ref{F1}. As each patch of the solution describes a flux of
  ingoing null radiation, this combined solution models in addition to an
  influx, a light-like particle which propagates towards positive infinity.
  The spacetime in region I is characterized by a mass function $m =
  m_{1}(v_{1})$ whereas region II is characterized by another
  mass function $m = m_{2}(v_{2})$. From the coordinate system
  (\ref{S2E2}), the outgoing null geodesic satisfies the equation
  \begin{eqnarray}
  2\,\dot{x}(\lambda) & = &
  \alpha\left(\,x(\lambda),v(\lambda)\,\right)\,\dot{v}(\lambda) \comma
  \label{S2E6}
  \end{eqnarray}
  where $\lambda$ is an affine parameter and
  the dot denotes derivative with respect to $\lambda$. Without loss of
  generality, we choose the parameter $\lambda$ to be zero
  at the Cauchy horizon and positive beyond that.

  \begin{figure}
    \leavevmode
    \hfil \boxit{\hbox{\epsfysize=8cm \epsffile{F_1.ps}}} \hfil
    \caption{A null line S divides the spacetime into regions I and II.}
    \label{F1}
  \end{figure}

  We now define a function $X$ to be the value of $x$ along $S$
  such that one of the Euler-Lagrange equations implies
  \begin{eqnarray}
    \ddot{v}(\lambda) & = &
    \minus \frac{1}{Q}\,(\dot{v})^2\,\left[\,{\cal M}'(X) - Q\,{\cal M}(X) +
Q\,m(v)\,\right]\,\EXP^{2\,\phi(X)} \label{S2E7}
  \end{eqnarray}
  along S. If we define
  \begin{eqnarray}
    z(\lambda) & := &
    \EXP^{\minus 2\,\phi(X(\lambda))} / \dot{v}(\lambda) \comma \label{S2E8}
  \end{eqnarray}
  one can show that
  \begin{eqnarray*}
    \dot{z}(\lambda) \b = \b \frac{1}{Q}\,{\cal M}'(X(\lambda)) & {\rm and} &
    m(v(\lambda)) \b = \b
    {\cal M}(X(\lambda)) - Q\,z(\lambda)\,\dot{X}(\lambda) \period
  \end{eqnarray*}
  As a result, we obtain the matching equations
  \begin{eqnarray}
    m(v(\lambda)) & = &
    {\cal M}(X(\lambda))
    - Q\,z(\lambda)\,\dot{X}(\lambda) \comma \label{S2E9A} \\
    v(\lambda) & = &
    \int^\lambda \EXP^{\minus 2\,\phi(X(\zeta))} / z(\zeta)\,\de \zeta \comma
\label{S2E9B} \\
    z(\lambda) & = &
    Z + \frac{1}{Q}\,\int^\lambda_0 {\cal M}'(X(\zeta))\,\de \zeta
\label{S2E9C}
  \end{eqnarray}
  if the boundary function $X$ is known. The term $Z$ in equation
  (\ref{S2E9C}) is an integration constant and we shall see that
  it plays an important role in the analysis.

  We place a subscript 1 to the terms $m$, $v$, $z$ and $Z$ in
  region I and subscript 2 to denote those terms in region II.
  Thus the ($\lambda$-dependent) ``mass'' of the null particle can be
  defined as
  \begin{eqnarray}
    \Delta m(\lambda) \b := \b m_{2}(\lambda) - m_{1}(\lambda) \b = \b
    Q\,(Z_{1} - Z_{2})\,\dot{X}(\lambda) \period \label{S2E10}
  \end{eqnarray}
  We also define a constant
  \begin{eqnarray}
    M := m_{1}(\lambda) + \delta m(\lambda) \label{S2E11}
  \end{eqnarray}
  as the final ADM mass of the black hole observed in region I after
  the hole has absorbed all the ingoing radiation. Therefore, the
  term $\delta m$ is interpreted as the mass of the radiative tail
  of the ingoing radiation.
  Since the Cauchy horizon corresponds to the limit
  $v_{1} \rightarrow \infty$, we expect that
  \begin{eqnarray*}
    \lim_{\lambda \rightarrow 0^{\minus}} \dot{v}_{1}(\lambda) \b = \b
    \EXP^{\minus 2\,\phi(X_{c})} / Z_{1} \b = \b \infty \period
  \end{eqnarray*}
  This implies that $Z_{1} = 0$, yielding
  \begin{eqnarray}
    \lim_{X \rightarrow X_c} {\cal M}(X) & = & M \label{S2E12}
  \end{eqnarray}
  by applying the limit to equation (\ref{S2E9A}).

  Equation (\ref{S2E9C}) can be written as
  \begin{eqnarray}
    z(\lambda) & = &
    Z - k_o\,\EXP^{\minus 2\,\phi(X_{c})}\,\lambda \comma \label{S2E13}
  \end{eqnarray}
  where $k_o$ is defined as
  \begin{eqnarray}
    k_o \b := \b
    \minus \frac{1}{Q}\,{\cal M}'(X(\epsilon))\,\EXP^{2\,\phi(X_{c})}
\label{S2E14}
  \end{eqnarray}
  by Mean Value Theorem where the small constant
  $\epsilon\,\in (\,\lambda,0\,)$. It is always possible to make
  $k_o$ to be positive definite if $|\,\lambda\,|$ is sufficiently small
  because the slope of ${\cal M}(X)$ at $X_c$ is non-positive. A
  sample graph of ${\cal M}(X)$ is shown in Figure~\ref{F2}. As a
  result, we have $k_o > 0$.

  \begin{figure}
    \leavevmode
    \flushleft \epsfysize=8cm \epsffile{F_2.ps}
    \caption{A sample graph of ${\cal M}(X)$.}
    \label{F2}
  \end{figure}

  With the help of equations (\ref{S2E11}) and (\ref{S2E12}),
  equation (\ref{S2E9A}) can be approximated and gives
  \begin{eqnarray}
    \dot{X}(\lambda) & \approx &
    \frac{1}{k_o\,Q\,\lambda}\,\delta m(\lambda)\,\EXP^{2\,\phi(X_{c})}
\label{S2E15}
  \end{eqnarray}
  when $|\,\lambda\,| \ll 1$. Since $Z_{1} = 0$,
  equation (\ref{S2E9B}) can be approximated as
  \begin{eqnarray}
    v_{1}(\lambda) \b = \b
    \minus \frac{1}{k_o}\,\int^\lambda \EXP^{2\,\phi(X_{c}) -
2\,\phi(X(\zeta))}\,\frac{\de \zeta}{\zeta}
    \b \approx \b \minus \frac{1}{k_o}\,\int^{\lambda} \frac{\de \zeta}{\zeta}
    \b = \b \minus \frac{1}{k_o}\,\ln|\,\lambda\,| && \label{S2E16}
  \end{eqnarray}
  in some negative neighborhood of $\lambda = 0$. However, when
  $|\,\lambda\,| \ll 1$, $z_{2} \approx Z_{2}$ which implies
  \begin{eqnarray}
    v_{2}(\lambda) \b \approx \b
    \int^{\lambda} \EXP^{\minus 2\,\phi(X_{c})} / Z_{2}\,\de \zeta
    \b = \b \EXP^{\minus 2\,\phi(X_{c})}\,\frac{\lambda}{Z_{2}} \period
\label{S2E17}
  \end{eqnarray}

  As $\delta m$ represents the radiative tail of the ingoing radiation,
  its relevant asymptotic behaviour is
  \begin{eqnarray}
    \delta m(v_{1}) & = & h\,|\,v_{1}\,|^{\minus p} \comma \label{S2E18}
  \end{eqnarray}
  where $h$ and $p$ are positive constants (see the appendix for a
  discussion on this point). When $\lambda$ is
  close to zero, we find that $\Delta m$ is approximately
  \begin{eqnarray}
    \Delta m(\lambda) \b = \b \minus Q\,Z_{2}\,\dot{X}(\lambda)
    \b \approx \b
    \minus
\frac{h\,Z_{2}}{k_o\,\lambda}\,\EXP^{2\,\phi(X_{c})}\,|\,v_{1}(\lambda)\,|^{\minus p} \period \label{S2E19}
  \end{eqnarray}
  As a result, we obtain
  \begin{eqnarray}
    \Delta m(v_{1}) & \approx &
    \frac{h\,Z_{2}}{k_o}\,\EXP^{2\,\phi(X_{c})}\,|\,v_{1}\,|^{\minus
p}\,\EXP^{k_o\,v_{1}} \period \label{S2E20}
  \end{eqnarray}
  On the other hand, when $|\,\lambda\,| \ll 1$, we have
  \begin{eqnarray}
    m_{2}(\lambda) & = &
    M - \delta m(\lambda) + \Delta m(\lambda) \nonumber \\
    m_{2}(v_{2}) & \approx &
    M - h\,\left[\,1 +
\frac{1}{k_o\,v_{2}}\,\right]\,|\,k_o\,|^p\,\left|\,\ln\left|\,Z_{2}\,\EXP^{2\,\phi(X_{c})}\,v_{2}\,\right|\,\right|^{\minus p} \period \label{S2E21}
  \end{eqnarray}
  This shows that the mass in region II becomes unbounded near the
  Cauchy horizon where $v_2 \approx 0$.

\bigskip

\section{The Second Solution} \label{Sct3}
  In the case when $\gamma = \sigma = 2$ and $Q = 0$, the action
  (\ref{S1E1}) yields a set of field equations which has a solution
  \begin{eqnarray}
    f & = & \frac{q}{(x - x_o)^2} \comma \label{S3E1A} \\
    \phi & = & \minus \ln|\,x - x_o\,| \comma \label{S3E1B} \\
    \alpha & = & \frac{{\cal M}(x) - m(v)}{x - x_o} \comma \label{S3E1C} \\
    {\cal M}(x) & = &
    \left(\,x - x_o\,\right)\,\left[\,1 + \frac{q^2}{4\,(x - x_o)^2} -
\half\,\sum^{k}_{n=2} \frac{a_n}{2\,n-3}\,(x - x_o)^{2-2\,n}\,\right]
\label{S3E1D} \B
  \end{eqnarray}
  in the Eddington-Finkelstein coordinate system (\ref{S2E2}) with the use
  of the same stress-energy tensor (\ref{S2E4}). The constants $q$
  and $x_o$ are again the integration constants and the function
  $m(v)$ which satisfies the same differential equation (\ref{S2E5})
  is the ADM mass of the black hole. As the singularity is located
  at $x=x_o$, the spacetime is divided into two disjoint sets:
  $\Re \times \{ (\minus \infty, x_o) \}$ and
  $\Re \times \{ (x_o, \infty) \}$. Thus the whole real line
  consists of two disjoint universes and each of them has a
  one-way-collapse solution (\ref{S3E1A}) to (\ref{S3E1D}).
  In static case, as the ingoing coordinate $v$ is defined as
  \begin{eqnarray*}
    v & = & t + \int \frac{\de x}{\alpha(x)} \comma
  \end{eqnarray*}
  where $t$ is the usual Schwarzschild coordinate time, it tends to
  negative (positive) infinity when $x$ in the left (right) universe
  approaches to the Cauchy horizon.

  Consider the matching of two patches of solution
  (\ref{S3E1A})--(\ref{S3E1D})
  along the outgoing thin null particle S as in the previous section.
  This time we define $z$ as
  \begin{eqnarray}
    z(\lambda) & := &
    \left[\,X(\lambda) - x_o\,\right] / \dot{v}(\lambda) \label{S3E2}
  \end{eqnarray}
  such that one of the Euler-Lagrange equations yields
  \begin{eqnarray}
    \ddot{v}(\lambda) & = &
    \minus \half\,(\dot{v})^2\,\left[\,\left(\,X - x_o\,\right){\cal M}'(X) -
{\cal M}(X) + m(v)\,\right]\,\left(\,X - x_o\,\right)^{\minus 2} \B
\label{S3E3}
  \end{eqnarray}
  along the null line S. Thus, the system analogous to (\ref{S2E9A})
  to (\ref{S2E9C}) is
  \begin{eqnarray}
    m(v(\lambda)) & = &
    {\cal M}(X(\lambda))
    - 2\,z(\lambda)\,\dot{X}(\lambda) \comma \label{S3E4A} \\
    v(\lambda) & = &
    \int^\lambda \left[\,X(\zeta) - x_o\,\right] / z(\zeta)\,\de \zeta \comma
\label{S3E4B} \\
    z(\lambda) & = &
    Z
    + \half\,\int^\lambda_0 {\cal M}'(X(\zeta))\,\de \zeta \period
\label{S3E4C}
  \end{eqnarray}
  Again we use the subscripts 1 and 2 to distinguish the quantities
  $m$, $v$, $z$ and $Z$ in region I and II.

  Making use of the same definition as in (\ref{S2E10}), we find
  \begin{eqnarray}
    \Delta m(\lambda) & = &
    2\,(Z_{1} - Z_{2})\,\dot{X}(\lambda) \period \label{S3E5}
  \end{eqnarray}
  As $Z_{1}$ is zero, it is clear that
  ${\rm sign}(Z_{2}) = {\rm sign}(X - x_o)$ in order to have
  positive $\Delta m$.
  Similar to the previous case, equation (\ref{S3E4C}) can be expressed as
  \begin{eqnarray}
    z_{1}(\lambda) & = &
    Z_{1} - k_o\,\left(\,X_{c}^{\pm} - x_o\,\right)\,\lambda \label{S3E6}
  \end{eqnarray}
  by the use of Mean Value Theorem, where $k_o$ is defined as
  \begin{eqnarray}
    k_o & := &
    \minus \half\,{\cal M}'(X(\epsilon)) / \left(\,X_{c}^{\pm} - x_o\,\right)
\label{S3E7}
  \end{eqnarray} and $\epsilon$ is in the open interval $(\lambda,0)$. The
constants
  $X_{c}^{\pm}$ are the locations of Cauchy horizons in the left and
  right universes about $x_o$. A sample ${\cal M}-X$ graph is shown
  in Figure~\ref{F3}. In considering the universe left (right)
  of $x_o$, one must choose $X_{c}^{-}$ ($X_{c}^{+}$) as the small $\lambda$
  approximation of $X$ in equation (\ref{S3E6}).
  By using the same argument as before, we have $k_o > 0 \b (< 0)$
  in the right (left) universe. Therefore in some small
  negative neighborhood of $\lambda = 0$, equation (\ref{S3E4A})
  can be approximated as
  \begin{eqnarray}
    \dot{X}(\lambda) & \approx &
    \frac{1}{2\,k_o\,\lambda}\,\delta m(\lambda) / \left|\,X_{c}^{\pm} -
x_o\,\right| \label{S3E8}
  \end{eqnarray}
  because $Z_{1}$ is zero.

  In region I, $v_{1}$ can be approximated as
  \begin{eqnarray}
    v_{1}(\lambda) \b = \b
    \minus \int^\lambda \frac{X(\zeta) - x_o}{X_{c}^{\pm} - x_o}\,\frac{\de
\zeta}{k_o\,\zeta} \b \approx \b
    \minus \int^\lambda \frac{\de \zeta}{k_o\,\zeta} \b = \b
    \minus \frac{1}{k_o}\,\ln|\,\lambda\,|\label{S3E9}
  \end{eqnarray}
  in some negative neighborhood of $\lambda = 0$. Nevertheless when
  $|\,\lambda\,| \ll 1$, $z_{2} \approx Z_{2}$ which implies that
  \begin{eqnarray}
    v_{2}(\lambda) \b \approx \b
    \frac{X_{c}^{\pm} - x_o}{Z_2}\,\lambda \period \label{S3E10}
  \end{eqnarray}
  Notice that equations (\ref{S3E9}) and (\ref{S3E10}) are similar
  to (\ref{S2E16}) and (\ref{S2E17}) in the last section and it is
  the logarithmic property of $v_{1}$ and linearity of $v_{2}$
  which trigger the mass inflation mechanism in region II.

  \begin{figure}
    \leavevmode
    \epsfxsize=1cm \epsfysize=8cm \epsffile{F_3.ps}
    \caption{A sample graph of ${\cal M}(X)$.}
    \label{F3}
  \end{figure}

  By using equation (\ref{S3E5}), we find that $\Delta m$
  can be approximated as
  \begin{eqnarray}
    \Delta m(\lambda) \b = \b \minus 2\,Z_{2}\,\dot{X}(\lambda)
    \b \approx \b
    \minus \frac{h\,Z_{2}}{k_o\,\left|\,X_{c}^{\pm} -
x_o\,\right|\,\lambda}\,|\,v_{1}(\lambda)\,|^{\minus p} \label{S3E11}
  \end{eqnarray}
  which can also be written as
  \begin{eqnarray}
    \Delta m(v_{1}) & \approx &
    \frac{h\,Z_{2}}{k_o\,\left|\,X_{c}^{\pm} -
x_o\,\right|}\,|\,v_{1}\,|^{\minus p}\,\EXP^{k_o\,v_{1}} \period \label{S3E12}
  \end{eqnarray}
  On the other hand, we can write
  \begin{eqnarray}
    m_{2}(v_{2}) & \approx &
    M - h\,\left[\,1 - \frac{1}{|\,k_o\,v_{2}\,|}\,\right]\,|\,k_o\,|^p\,
    \left|\,\ln\left|\,\frac{Z_2\,v_{2}}{X_{c}^{\pm} - x_o}\,\right|\,
    \right|^{\minus p} \label{S3E13}
  \end{eqnarray}
  and so the mass in region II in this case is also unbounded near
  the Cauchy horizon.

  Finally, when we calculate the dyad component $R^1_{\ 001}$ of the
  Riemann tensor using either the metric (\ref{S2E3C}) or (\ref{S3E1C}),
  we find that the leading term of this component has the form
  \begin{eqnarray}
    R^1_{\ 001} & \approx &
    {\rm constant} \times \left|\,\ln|\,v_{2}\,|\,\right|^{\minus p}/v_{2}^3
\label{S3E14}
  \end{eqnarray}
  which is exactly the same as the behaviour for the corresponding
  component associated with the $(3+1)$-dimensional
  Reissner-Nordstr\"{o}m metric. Hence the
  tidal distortion near the Cauchy horizon is also bounded in both
  solutions even though Ricci scalar is infinite there.


\bigskip

\section{Conclusions} \label{Sct4}

We have shown that $(1+1)$-dimensional charged dilatonic black holes have yet
another feature in common with their $(3+1)$-dimensional counterparts,
namely that they mass inflate as a consequence of the interaction between
incoming and backscattered radiation. This property holds even if the black
holes have multiple horizons (as in eq. (\ref{S2E3D})).   Moreover, the
singularity associated with it is also as mild as the one in $(3+1)$
dimensions. Since the quantization of lower dimensional gravity is
considerably easier than the $(3+1)$-dimensional case, it should be
possible to extend the results of this paper to include quantum mechanical
effects. Work on this is in progress.


\section*{Acknowledgements}
  After completion of this paper, we became aware of related work by S. Droz
  on mass-inflation in $(1+1)$ dimensions. We wish to thank Dr. V. Husain
  for drawing this to our attention. This work was supported in part by the
  Natural Sciences and Engineering Research Council of Canada.

\appendix
\section{Scalar perturbation in $(1+1)$ dimensions} \label{ApdA}
  In $(1+1)$ dimensions, the usual scalar wave equation
  $\Del^2 \psi = 0$ is not appropriate to be used in the perturbation
  problem because there is no potential barrier and hence there is
  no backscattering \cite{Price}. As a result, we use an alternative
  wave equation
  \begin{eqnarray}
    \Del^2 \psi - \xi\,R\,\psi & = & 0 \comma \label{A1E1}
  \end{eqnarray}
  where $\xi$ is a constant, to describe a radiating scalar field.
  If we put the wave equation into a conformally flat coordinate system
  \begin{eqnarray}
    \de s^2 & = &
    \hat{\alpha}(X)\,\left[\,\minus \de T^2 + \de X^2\,\right]
    \b = \b 2\,\de x\,\de v - \alpha(x)\,\de v^2 \comma \label{A1E2a}
  \end{eqnarray}
  where $\alpha(x)$ is taken to be
  \begin{eqnarray}
    \alpha(x) & = & 1 - \frac{M}{x} \label{A1E3}
  \end{eqnarray}
  for simplicity, the wave equation (\ref{A1E1}) will become
  \begin{eqnarray}
    \wave \psi & = & B(X)\,\psi \comma \label{A1E2b}
  \end{eqnarray}
  where the potential barrier $B(X)$ is given by
  \begin{eqnarray}
    B(X) & = &
    \frac{4\,\xi}{M^2}\,\Omega\left(\,\frac{1}{M}\,\EXP^{(X-M) /
M}\,\right)\,\left[\,1 + \Omega\left(\,\frac{1}{M}\,\EXP^{(X-M) /
M}\,\right)\,\right]^{\minus 4} \label{A1E2c}
  \end{eqnarray}
  and the function $\Omega$ is defined implicitly as
  $\Omega(x)\,\exp(\Omega(x)) = x$. Figure~\ref{F4} is the graph
  of this barrier and one can see that it has a peak at about
  $X = M + M\,\ln(M / 2)$.

  \begin{figure}
    \leavevmode
    \epsfysize=8cm \epsffile{F_4.ps}
    \caption{The graph of the potential barrier $B(X)$.}
    \label{F4}
  \end{figure}

  Let us concentrate on a static background in an outgoing coordinate system
  \begin{eqnarray*}
    \de s^2 & = &
    \minus 2\,\de x\,\de u - \alpha(x)\,\de u^2 \period
  \end{eqnarray*}
  In this coordinate system, our wave equation
  (\ref{A1E1}) has a form
  \begin{eqnarray}
    \alpha(x)\,\dixx \psi - 2\,\partial_{xu} \psi + \alpha'(x)\,\dix \psi
    + \xi\,\alpha''(x)\,\psi & = & 0 \period \label{A1E4}
  \end{eqnarray}
  As the potential barrier of (\ref{A1E1}) is of order  $1 / x^3$,
  one will expect that the resultant backscattering in this spacetime
  is similar to the one in $(3+1)$ dimensions in which the leading
  order of the potential barrier is of the order $1 / r^3$. This
  can be verified mathematically as follows.

  In Kruskal coordinates, the scalar field is expected to be
  well-behaved at the event horizon because the coordinates are
  well-behaved there. As $U := \minus \exp( \minus u / 2\,M )$ and
  $V := \exp( v / 2\,M )$, we have
  \begin{eqnarray*}
    \frac{\de \psi}{\de u} & = &
    \frac{\di \psi}{\di U}\,\frac{\de U}{\de u}
    + \frac{\di \psi}{\di V}\,\frac{\de V}{\de u} \b = \b
    \frac{1}{2\,M}\,\frac{\di \psi}{\di U}\,\EXP^{\minus u / 2\,M} \period
  \end{eqnarray*}
  As a result, we expect that
  \begin{eqnarray*}
    \psi \rightarrow a + b\,\EXP^{\minus u / 2\,M} & {\rm as} &
    t \rightarrow \infty \B ({\rm i.e.} \B u \rightarrow \infty) \period
  \end{eqnarray*}
  Thus the primary outgoing waves in $(1+1)$ dimensions decay to zero
  exponentially as those in $(3+1)$ dimensions.

  Since the primary outgoing waves have the form
  \begin{eqnarray}
  \psi(x,u) & = & \sum^\infty_{n=1} \frac{f_n(u)}{x^n} \comma \label{A1E5}
  \end{eqnarray}
  a difference equation
  \begin{eqnarray}
  f_n'(u) & = &
  \minus \frac{n-1}{2}\,f_{n-1}(u)
  + \left(\,\frac{n-2}{2} + \frac{\xi}{n}\,\right)\,M\,f_{n-2}(u) \label{A1E6}
  \end{eqnarray}
  can be obtained by using equations (\ref{A1E4}) and (\ref{A1E5}).
  This difference equation is the same as the one below part 1(b)
  in Box 3, reference \cite{Thorne} when $l = 0$ and $B_n = 0$
  except $B_1 = \xi$. One can therefore follow the rest of the calculation
  in \cite{Thorne} to show that the backscattered waves die out as
  $t^{\minus p}$.

  When $\alpha(x)$ has more than one horizon, the dominant part of
  the metric is still given by (\ref{A1E3})
  because we are considering the backscattering which takes place just
  outside of the event horizon \cite{Price} where $x$ is still large.
  Thus the $1 / x$ term dominates the terms $1 / x^2$, $1 / x^4$ and so on.
  As a result, we still expect that the
  backscattered waves decay as $t^{\minus p}$. Finally, in the case
  when the spatial function in the metric is exponential rather than
  inverse of $x$, the potential barrier is thinner than in the case
  considered above
  because the barrier in the exponential falloff is not as
  widespread as the barrier in the case of inverse decay.
  Thus the decay of the backscattered waves is even faster
  than $t^{\minus p}$ because more waves can tunnel through the
  barrier and propagate to infinity.

\end{document}